\documentstyle[aps,epsfig,multicol]{revtex}
\newcommand{\beq}{\begin{equation}}
\newcommand{\eeq}{\end{equation}}
\newcommand{\bea}{\begin{eqnarray}}
\newcommand{\eea}{\end{eqnarray}}
\newcommand{\eps}{\varepsilon}
\begin{document}
%\draft
%\widetext
\title{Semiclassical cross section correlations}
\author{Bruno Eckhardt$^1$, Shmuel Fishman$^2$ and Imre Varga$^{1,3}$}
\address{$^1$Fachbereich Physik, Philipps Universit\"at Marburg,
D-35032 Marburg, Germany}
\address{$^2$Department of Physics, Technion,
Haifa 32000, Israel}
\address{$^3$Elm\'eleti Fizika Tansz\'ek, Fizikai Int\'ezet,
Budapesti M\H uszaki \'es Gazdas\'agtudom\'anyi Egyetem,
H-1524 Budapest, Hungary}
\date{\today}
\maketitle

\begin{abstract}
We calculate within a semiclassical approximation the autocorrelation
function of cross sections. The starting point is the
semiclassical expression for the diagonal matrix elements of
an operator. For general operators with a smooth classical limit
the autocorrelation function of such matrix elements has two
contributions with relative weights determined by classical
dynamics. We show how the random matrix result can be
obtained if the operator approaches a projector onto
a single initial state.
The expressions are verified in calculations for the
kicked rotor.
\end{abstract}


\begin{multicols}{2}
\section{Introduction}
Quantum systems whose classical limit is chaotic show fluctuations
in cross sections and eigenvalue positions whose statistical
properties seem to fall into a few universality classes
\cite{Haake,Bohigas,Guhr}.
Among the many measures that have been applied to characterize
these statistical features,  much attention has been
given to two-point correlation functions since they can
under certain assumptions be related to the classical
dynamics \cite{Berry85}. For the case of spectra of bounded systems
this has worked remarkably well and in addition one
of the main predictions of the semiclassical analysis, the existence
of long range correlations due to periodic orbits \cite{Berry85,Wintgen85},
has been confirmed many times \cite{Haake,Stoeckmann}.

More recently investigations of the statistical behaviour of
directly observable quantities, such as cross sections, have been
worked out within the nonlinear sigma model for disordered systems
\cite{Alhassid,Alhassid2}. The correlation function was found to have
two contributions, a Lorentzian and a derivative of a Lorentzian with
respect to its parameter. The ratio between the two terms is fixed and
depends on symmetry only.
Since there is no semiclassical expression
for the individual wave functions from which the
cross sections could be calculated, the derivation of
such correlation functions within semiclassics poses a
serious challenge. A first step in this direction
was undertaken by Agam \cite{Agam}, who exploits
quantum properties of the matrix elements and does
not use previously established formulas for diagonal
matrix elements \cite{Wilkinson,Eckh92b}. The
derivation presented here is similar in spirit but
starts from the semiclassical expression
for diagonal matrix elements and specializes to the case of
the cross section in the end. In particular,
we show how the relative weight between the two contributions
to the correlation function can be changed.
The final expressions are compared with data for
cross sections in an open kicked rotor model.

In section 2 we present the semiclassical derivation of the
correlation function between cross sections.
This calculation is actually straightforward and closely
patterned after calculations for other two point
correlations~\cite{Berry85}.
In section 3 we discuss the limit that has to be
taken in the observable to arrive at the correlation
function for cross sections. In section 4 we discuss
numerical simulations for an open kicked rotator.
Some concluding comments are given in section 5.

\section{Semiclassical correlation functions for smooth operators}
Quantum cross sections for the transition from an initial state
$|i\rangle$ to a set of final states $|n\rangle$ which are eigenstates
of a Hamiltonian, $H|n\rangle=E_n |n\rangle$, are proportional
to
\beq
\sigma(E) \propto
\sum_n \langle n|D|i\rangle \langle i|D |n\rangle \delta(E-H)
\eeq
where $D$ denotes the dipole operator. Using the projection
operator
\beq
A = D|i\rangle \langle i|D
\eeq
and Greens function $G=1/(E-H)$ the cross section becomes (dropping the
proportionality factors implied in (1))
\beq
\sigma_A(E) = -{1\over\pi} \mbox{Im\ tr} (G\, A) \,.
\eeq
If the operator $A$ is sufficiently smooth and has a non-singular
classical limit, there is a semiclassical expression for
$\sigma(E)$ which naturally divides into two pieces
\cite{Wilkinson,Eckh92b},
\beq
\sigma^{(sc)}_A(E) = \sigma_0(E) + \sigma_{A,fl}(E)\,.
\eeq
The first term,
a smoothly varying background contribution from paths of
`zero length', is determined by integration over the energy
shell of the Wigner transform $A_W$ of the observable with the
measure $d\mu = d^dp d^dq / h^d$ (in $d$-degrees of freedom),
\beq
\sigma_0(E) = \int d\mu A_W \delta(E-H) \,.
\label{sigma0}
\eeq
The second term describes the fluctuations around it and is
determined by classical periodic orbits,
\beq
\sigma_{A,fl}(E) =
\frac{1}{\pi\hbar}
\mbox{Re\ } \sum_p A_p w_p e^{iS_p/\hbar}
\eeq
where $S_p$, $T_p$ and $w_p$ are the action, period and weight of the
periodic orbit $p$, respectively, and
\beq
A_p = \int_0^{T_p} dt\, A_W(p(t),q(t))
\eeq
is the integral of the observable (again the Wigner transform
of $A$) over the periodic orbit $p$.
This result is derived under the condition that the observable is
sufficiently smooth so that it does not affect the stationary
phase evaluation that singles out periodic orbits.
Wigner transformations of projection operators are critical
in this respect since they typically approach delta functions
in momenta as $\hbar$ goes to zero, i.e. become rather
singular. This is the main limitation that prevents
a direct application of the above expression to the calculation
of the autocorrelation function of the cross sections.
We therefore adopt the following strategy: We first calculate
the auto-correlation function for $\sigma_A(E)$ within a semiclassical
approximation for a smooth observable $A$ and then discuss
the limit of a singular operator.

The object we want to calculate is the normalized
autocorrelation function
of the fluctuations around the mean cross section $\sigma_0(E)$,
\beq
C(\eps) = \langle \sigma_{A,fl}(E+\eps/2) \sigma_{A,fl}(E-\eps/2) \rangle
/\langle \sigma_0\rangle^2\,
\eeq
where $\langle\ldots\rangle$ denotes an average over energy.
The energy scale is set quantum mechanically by the mean
spacing $\Delta$ between neighboring levels, calculated from
the mean density of states
\beq
\rho_0=1/\Delta = \int d\mu \delta(E-H)\,.
\eeq
The associated
time scale is the Heisenberg time $T_H = 2\pi\hbar \rho_0 = h/\Delta$.
The scale for the observable is set by the phase space
average of its Wigner transform,
\beq
\overline{A} = {\int d\mu A_W \delta(E-H) \over \int d\mu \delta(E-H)}
\eeq
so that $\sigma_0 = \overline{A} \rho_0 = \overline{A} / \Delta$.

Substituting the fluctuating part from the periodic orbits gives
\bea
    C(\eps) &=& \frac{\Delta^2}{2\overline{A}^2\pi^2\hbar^2} \times
    \nonumber\cr
            &\ &
    \mbox{Re} \left\langle
    \sum_{p,p'} A_p^* A_{p'}  w_p^* w_{p'}
    e^{i(S_p(E+\eps/2)-S_{p'}(E-\eps/2))/\hbar} \right\rangle\,.
\eea
The classical action in the exponent can be expanded for small $\eps$,
\bea
S_p(E+\eps/2) &-& S_{p'}(E-\eps/2) \approx \nonumber\cr
              &\ & S_p(E)-S_{p'}(E) + \eps (T_p+T_{p'})/2 \,.
\eea
In the diagonal approximation \cite{Berry85}
the correlation function becomes
\beq
    C_{diag}(\eps) =\frac{g\Delta^2}{2\bar{A}^2\pi^2\hbar^2} \mbox{Re}
    \sum_{p}|A_p|^2 |w_p|^2 e^{iT_p \eps/\hbar} \,
\eeq
the factor $g$ accounts, as usual, for degeneracies in case of real
symmetric Hamiltonians, $g=2$, while for hermitian Hamiltonians $g=1$.

For the next step we use the periodic orbit sum rule
\cite{HOdA} and follow the steps in \cite{Eckh95}.
The periodic orbits proliferate exponentially with time so that
the sum on $p$ can be replaced by an integral over time.
The density of orbits is given by $e^{\mu T}/T$
with $\mu$ the topological entropy
and their weight by $|w_p|^2 = e^{-\lambda T}$ with $\lambda$
the Lyapunov exponent. The difference between
the Lyapunov exponent and the topological entropy is the
classical escape rate,
\beq
\Gamma = \lambda - \mu \,.
\eeq
The integrals $A_p$ of the observable along the orbit vary considerably
among orbits of similar length. The high density of periodic orbits
allows to capture this probabilistically through the distribution
$P(A)$
of values obtained for all orbits with periods in a small interval
around $T$. If the correlations in
the classical dynamics fall off sufficiently rapidly,
the distribution will be Gaussian,
\beq
P(A) = \frac{1}{\sqrt{\pi}s_A} e^{-(A-\tilde{A})^2/s_A^2}
\eeq
with a mean $\tilde{A} = \bar{A} T$ following from ergodicity and a
variance $s^2_A=\alpha T$ that increases linearly with time.
With this distribution function and the assumption
that there are no correlations between
weights $w_p$ and observables $A_p$, the mean square average
of the $A_p$'s from orbits with periods near $T$
changes with time like
\beq
\langle |A_p|^2 \rangle (T) = \bar{A}^2 T^2 +
\alpha T \,.
\label{Asquare}
\eeq
Thus summing the contributions of orbits in the
diagonal approximation we obtain
\bea
C_{diag}(\eps) &=& \frac{g\Delta^2}{2\bar{A}^2\pi^2\hbar^2}\times
 \nonumber\cr
               &\ &
 \mbox{Re}
    \int_0^{T_H} dT  \frac{e^{(\mu-\lambda+i\eps/\hbar) T}}{T}
\left( \overline{A}^2 T^2 +\alpha T\right) \,.
\eea
For bounded systems ($\mu=\lambda$) and for $\epsilon=0$ we obtain
the expression for the variance of matrix elements derived in 
\cite{Eckh95} and tested and verified in many situations 
\cite{Mehlig,Keating,Baecker}.
If the system is sufficiently open so that
$\Gamma T_H\gg 1$ the contributions from the off-diagonal terms
can be neglected and the integration continued to infinity (as explained
in the next paragraph). Then
\beq
C_{diag}(\eps) \approx
  \frac{g\Delta^2}{2\overline{A}^2\pi^2\hbar^2} \mbox{Re\ }
\left(\alpha  \frac{1}{\Gamma-i\eps/\hbar} +
\overline{A}^2  \frac{1}{(\Gamma-i\eps/\hbar)^2} \right) \,.
\eeq
With the energy and time scales mentioned
above, i.e. $\eps = \tilde\eps \Delta$ and
$\Gamma = 2\pi \tilde\Gamma / T_H = \tilde\Gamma \Delta /\hbar$,
the correlation function becomes
\beq
C(\eps) \approx \frac{g}{\pi}\left(
\frac{\alpha}{\overline{A}^2 T_H}
\frac{\tilde\Gamma}{\tilde\Gamma^2+\tilde\eps^2} +
\frac{1}{2\pi}
\frac{\tilde\Gamma^2-\tilde\eps^2}{(\tilde\Gamma^2+\tilde\eps^2)^2}
\right)
\,.
\label{final}
\eeq
As in the calculation of Fyodorov and Alhassid \cite{Alhassid}
the correlation function has two terms, a Lorentzian and
a derivative of a Lorentzian with respect to the width.
However, in contrast to their formula, where the relative weight between
the two terms was fixed, it here depends on the observable and
the Heisenberg time. This point will be taken up again in the next section.

The above derivation is based on the usual assumptions on the diagonal
approximation, the validity of a periodic orbit sum rule and the
replacement of a sum over orbits by an integral in time.
As a consequence deviations can be expected for short times where isolated 
periodic orbits dominate, an effect that should be particularly noticeable 
near bifurcations. Deviations from the diagonal approximations are
strong for the orthogonal ensemble and absent for the unitary ensemble,
at least up to the Heisenberg time \cite{FK}. 
The extension of the time integration
up to infinity rather than the Heisenberg time is justified if
the system is very open, i.e. if $\Gamma T_H\gg 1$, so that the
corrections due to off-diagonal terms for times beyond the Heisenberg time
can be neglected. When approaching bounded systems
$\Gamma$ vanishes and the corrections have to be taken
into account. The final expression is thus reasonable
only for sufficiently open systems where $\Gamma/\hbar\gg\Delta$.
In this limit the contributions from long orbits are
quickly suppressed and the differences between the unitary
and orthogonal ensembles should disappear, except of course for the
factor of $2$.
It is possible to go beyond this by assuming
that the form factor is the random matrix form factor times
an exponential damping, as suggested also by
Alhassid and Fyodorov \cite{Alhassid2,Dresden}.

The derivative of the Lorentzian in the second term,
weighted by the mean of the operator can be traced back to the
autocorrelation function of the density of states (without operator)
as calculated earlier \cite{Eckh92}.
The width $\Gamma$ that enters here is the classical escape rate,
since that is what determines the modification of the classical
sum rule. The prediction then is that the quantum resonances have
half that width, since it is the probabilities and not the
amplitudes that have to follow the classical behaviour. 
%%%%% changed
In many
cases, especially with a finite number of channels, the situation
presumably is more complex, since the quantum resonances have
a distribution of widths and it is not clear which quantity
(average width, maximum of distribution, longest life time
etc) dominates the form factor \cite{Borgonovi}. 
%%%%% added
In the semiclassical limit of an infinite number of channels
some information can be drawn from the results of Fyodorov and
Sommers for random matrix models with fixed transmission \cite{Sommers} and
from the distribution of resonance widths calculated by
Haake {\it et al.} \cite{Haake2}: in both cases a single width parameter,
given by half the classical escape rate in the first case and
by half the gap in the distribution in the second case
suffices to describe the correlation function.

Finally, we note that if the classical correlations do not decay sufficiently
rapidly the average of the square of the integrals along the orbits can be 
expressed as an integral over the correlation function as in the previous
calculation of matrix elements \cite{Eckh95}.

\section{The limit of projection operators}
The final result for the correlation function given in Eq.
(\ref{final}) has the two functional dependencies also
identified in Alhassid and Fyodorov \cite{Alhassid},
but the relative weighting depends on the observable.
Even worse, the first term contains $T_H$ in the denominator
and therefore seems to vanish in the semiclassical limit
where $T_H$ diverges. Then the correlation function
is of the form of a derivative of the Lorentzian only.
So how can one obtain Alhassid and Fyodorov's result
$\alpha/\overline{A}^2T_H=1$ within this semiclassical approach?

The key to the problem is the observation that the
semiclassical approximation assumes the observable to
have a nonsingular classical limit whereas the quantum
cross sections are obtained from observables that
are projectors on the initial state (weighted with
the dipole operator). The Wigner transform of a
projector is itself a function of Planck's constant
and becomes singular in the semiclassical limit.
More specifically, one can visualize the Wigner
transform of a projector as a characteristic
function that in $d$ dimensions lives on a phase space cell of
volume $h^d$, since that is the phase space volume
occupied by a single state.
The observables for which the semiclassical trace formula
was derived were smooth with a non-singular limit, that is
to say they covered an increasing number of quantum states
in the semiclassical limit. This smearing over many states
suppresses the first term in the correlation function.

It is possible to estimate the consequences of this observation
on the operator $A$ in a simple model of a uniformly
damped quantum map. Consider a 2-d chaotic map
on a finite phase space and its quantum representation by
an $N\times N$ unitary operator $U$. The dimension $N$
of $U$, Planck's constant $h$ and the volume $\Omega_0$ of
the classical phase space are connected by $h = \Omega_0/N$.
Damping is introduced uniformly everywhere in phase space
and on all quantum states.
The projector is modelled by an observable that takes on
the value $a_0$ in some part of phase space of area $\Omega_A$
and vanishes everywhere else.

On the classical side the integrals of the observable along
the periodic orbit are replaced by sums over the points of the orbit.
If there are no correlations between different time steps,
the average value of the $A_p$ over all orbits with period $n$
is given by $\langle A_p\rangle = n a_0 p$, where $p=\Omega_A/
\Omega_0$ is the probability for a randomly chosen point
to lie in the phase space area where the observable does
not vanish. The second moment of the distribution is
given by
\beq
\langle A_p^2\rangle = a_0^2 p^2\, n^2 + a_0^2 p (1-p)\, n\,.
\eeq
By comparison with Eq.~(\ref{Asquare}) we read off
$\overline{A} = a_0 p$
and $\alpha=a_0^2 p (1-p)$. The Heisenberg time is $T_H=N$,
the dimension of the Hilbert space. Therefore,
\beq
{\alpha\over \overline{A}^2 T_H} = {a_0^2 p (1-p) \over a_0^2 p^2 N} =
{1-p \over pN}
\,.
\eeq
With an escape rate $\tilde\Gamma$ expressed in units of the
Heisenberg time the correlation function then becomes
\beq
C(\eps) =\frac{g}{\pi}
\left(
\frac{1- p}{p\, N}
\frac{\tilde\Gamma}{\tilde\Gamma^2+\eps^2} +  \frac{1}{2\pi}
\frac{\tilde\Gamma^2-\eps^2}{(\tilde\Gamma^2+\eps^2)^2}
\right)\,.
\eeq
This expression clearly shows the suppression of the first term
in the semiclassical limit of large $N$ if $p$ is fixed.

However, if the observable is a projector onto a single state
its Wigner transform should localize on a cell of
phase space volume $h$. Thus, $\Omega_A = \Omega_0/N$ and
$p=1/N$, so that the product $pN=1$. Except for the
tiny correction $1/N$ to the correct ratio of $1$
this is the result obtained by Alhassid and Fyodorov
\cite{Alhassid}.
In addition, it suggests a way to modify the
relative weighting between the two terms: consider
transitions not from a single state but from
$M$ states in an incoherent superposition, i.e.,
\beq
A = D \sum_m |m\rangle\langle m| \, D\,.
\label{Aproj}
\eeq
Then the phase space area covered by $A$ will increase
$M$-fold and the weight of the Lorentzian will
decrease correspondingly,
\beq
C(\eps) \approx \frac{g}{\pi}
\left(
\frac{1}{M}
\frac{\tilde\Gamma}{\tilde\Gamma^2+\eps^2} +  \frac{1}{2\pi}
\frac{\tilde\Gamma^2-\eps^2}{(\tilde\Gamma^2+\eps^2)^2}
\right)\,.
\label{ceps}
\eeq
(where it is assumed that $M \ll N$).
The resolution to the problem of the relative weight between
the two terms in the correlation function posed at the
beginning of this section
thus is that by focusing on initial states which
are projectors the classical observable also depends on
$\hbar$, and this $\hbar$-dependence influences the classical
quantities as well. Indeed, if the classical observable becomes very
localized in phase space, it is rarely visited, the
return time becomes large and the variance increases much more
slowly, on a timescale also set by the return time. In this way
the Heisenberg time enters the classical
quantities.

\section{Numerical tests on the kicked rotor}
The model of the previous section clearly
stretches the applicability of the
semiclassical expressions for the matrix elements
to their limits and requires numerical tests. We use
the kicked rotor in the momentum space
quantization of Izrailev \cite{Izraelev} for this purpose,
\beq
U_{nm}=\frac{1}{N}\sum_{l=0}^{N-1}e^{-iKN V(2\pi l/N)}
%\cos(2\pi l/N)}
e^{-2i\pi l(n-m)/N} e^{-i2\pi m^2/N} \, ,
\eeq
where \cite{Dresden} $V(\phi )= \cos\phi - \sin (2\phi)$
is the kicking potential. This model is known to belong to the
unitary universality class ($g=1$) due to the second term in the
potential that breaks the conjugation symmetry~\cite{BS,Maribor}.
If the kicking strength is sufficiently
large the correlations decay very quickly and one ends up with
essentially random values for the momenta. The calculations
were done for $K=7$~\cite{shepel} and matrix sizes $N=101$,
$201$, $401$, $801$ and $1601$.
The initial states were taken to be momentum eigenstates
of the unperturbed map. From the eigenstates $|\nu\rangle$ and
eigenphases $\phi_\nu$
a cross section was formed with the operator $A$ from
Eq.~(\ref{Aproj}) according to
\beq
\sigma(\phi) = \mbox{Re} \sum_\nu
{\langle \nu | A | \nu \rangle \over
1 - \exp(i(\phi-\phi_\nu)-\Gamma/2)}\,.
\label{dos}
\eeq
The damping $\Gamma/2$ is uniform for all eigenstates
and models the coupling of the system to a continuum
%%%% changed
(for a discussion of this point see the previous section
and the references cited there).
% That all states have the same decay rate is not very
% realistic but helps to avoid problems due to a
% variability in the damping constants and may \cite{Borgonovi}.
%%%%%
The role of the energy is now taken over by the phase
$\phi$ and the mean and correlation function are calculated
over the periodicity interval $2\pi$. The mean separation
between eigenphases is $\Delta=2\pi/N$.
Figure 1 shows the correlation function obtained
for a different number of initial states and constant
damping. For a single initial state the contribution from the 
derivative of a Lorentzian is barely noticeable but as the 
number of initial states increases the deviations from the
Lorentzian become larger.
In particular, the correlation function develops a
zero for $M/2\pi>\tilde\Gamma$. 
%%%%% this sentence is messed up:
Fitting Eq. (\ref{final}) with
the coefficient of the Lorentzian and $\tilde\Gamma$ as free
parameters yields a broadening $\tilde\Gamma\approx\Gamma$ 
independent of $M$
%%%%% and now what:  AND or ?
and the coefficient of the Lorentzian in Eq. 
(\ref{final}) is proportional to $M^{-1}$ as expected from 
Eq. (\ref{ceps}).

A random matrix calculation \cite{Maribor} shows that
the quantity that controls the relative weight between the two
terms is the ratio between the variance $s_A^2$ of the matrix elements
and the square of the average for different matrix sizes. This
quantity, $(s_A/\langle A\rangle)^2$ should equal $(1-p)/pN$.
As shown in Figure 2, the renormalized quantity
$(s_A/\langle A\rangle)^2pN$ follows the expected
$1-p$ behaviour rather closely, albeit with large fluctuations.
It thus seems that the assumptions that enter in our semiclassical
derivation of the correlation function can be satisfied in
chaotic systems.

\section{Final remarks}
We have shown how within a semiclassical approximation
correlation functions for cross sections in open
systems can be calculated. The calculations could
be supported with simulations in the standard map
and in particular the changes in the relative weight
between the two contributions to the correlation
function could be demonstrated.

Actually, the calculation works much better than
can reasonably be expected:
it is well known that the calculation of wave function
usually requires in addition to periodic orbits also
recurrent orbits \cite{Bogomolny,Huepper}. Their
importance depends on the width of the initial
state \cite{Eckh92b}. It may happen, however, that in
a statistical sense the differences between
the contributions from recurrent and periodic orbits
cancel. A related problem concerns the higher moments
of the distribution and thus the form of the full
distribution. Even in the singular limit of small
$p$ considered in the model, the distribution of
classical contributions remains Gaussian, or perhaps
Poissonian\cite{Maribor}, but the distribution for transition
strengths expected from random matrix theory is
exponential (for the unitary ensemble) or
Porter-Thomas (for the orthogonal ensemble).
Further calculations indeed show that while the first
and second moments agree, the higher moments and the full
distributions differ \cite{Maribor}.

Among the consequences that seem worthwhile to pursue are
the dependence on the initial state and the possibility
to highlight the non-Lorenzian part, the modification to
allow for non-exponential classical escape \cite{Dresden,Maribor}
and the singular contributions from periodic orbits
near bifurcations~\cite{Hamburg}.

\section*{Acknowledgments}
We thank O. Agam, Y. Alhassid and Y. Fyodorov for discussions.
This work was supported in part by the Alexander von Humboldt foundation
the Minerva Center for Non--linear Physics of Complex Systems and
by OTKA grants No. T029813, T032116 and F024135.

%%%%%%%%%%%%%%%%%%%%%%%%%%%%%%%%%%%%%%%%%%%%%%%%%%%%%%%%%%
% References
%%%%%%%%%%%%%%%%%%%%%%%%%%%%%%%%%%%%%%%%%%%%%%%%%%%%%%%%%%

%%%%%%%%%%%%%%%%%%%%%%%%%%%%%%%%%%%%%%%%%%%%%%%%%%%%%%%%%%%%%%%%%%%
% Figures
%%%%%%%%%%%%%%%%%%%%%%%%%%%%%%%%%%%%%%%%%%%%%%%%%%%%%%%%%%%%%%%%%%%%

\begin{figure}
\vspace{-0.3in}
\epsfxsize=8cm
\epsfbox{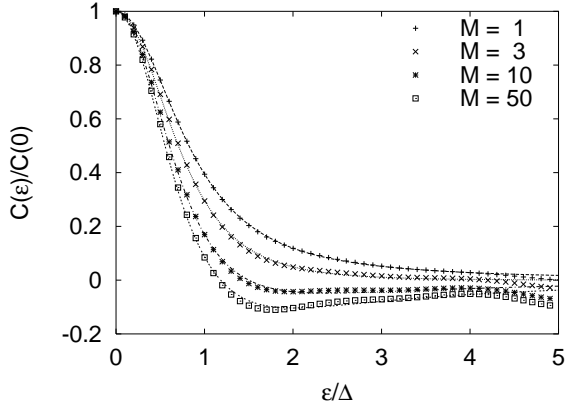}
\vspace{0.5cm}
\narrowtext
\caption[]{Cross section correlation function for different widths of
the observable, increasing from $M=1$ (top curve) to $3$, $10$
and $50$ (bottom curve). The width of the resonances was uniform throughout
the quantum spectrum: $\Gamma=\Delta$ (see Eq. (\protect\ref{dos})).
The continuous curves are fits to the functional form of 
Eq.~(\protect\ref{final}) with $\tilde\Gamma$ and the coefficient of the 
Lorentzian as fitting parameters with the condition $C(0)=1$. 
}\label{correl}
\end{figure}

\begin{figure}
\vspace{-0.3in}
\epsfxsize=8cm
\epsfbox{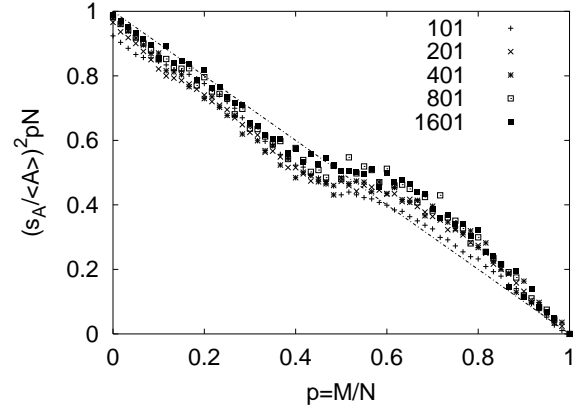}
\vspace{0.5cm}
\narrowtext
\caption[]{
Ratio between variance and average squared for observables
of different widths. The different symbols correspond to different
sizes of the matrix and thus to different values of $\hbar$.
}\label{variance}
\end{figure}

\end{multicols}

\begin{references}
\bibitem{Haake}F. Haake, Quantum Signatures of Chaos,
Springer, Berlin (1991);

\bibitem{Bohigas}
O. Bohigas, Chaos and Quantum Physics, Les Houches Lecture Notes,
Session LII, M.J. Giannoni, A. Voros and J. Zinn-Justin (eds),
North Holland, Amsterdam, 1991, p. 87;

\bibitem{Guhr} T. Guhr, A. M\"uller-Gr\"ohling, H.A. Weidenm\"uller,
Phys. Rep. {\bf 299}, 189 (1999).

\bibitem{Berry85} M.V. Berry, Proc. R. Soc. (London)
{\bf A400}, 229 (1985).

\bibitem{Wintgen85} D. Wintgen, Phys. Rev. Lett. {\bf 58}, 1589 (1987).

\bibitem{Stoeckmann} H.J. St\"ockmann, Quantum chaos: an introduction,
Cambridge University Press, 1999.

\bibitem{Alhassid}Y.V. Fyodorov and Y. Alhassid,
Phys. Rev. {\bf A58}, R3375 (1998).
% Y. Alhassid and Y. Fyodorov, cond-mat/9802105 and cond-mat/9808003

\bibitem{Alhassid2} Y. Alhassid and Y. V. Fyodorov,
J. Phys. Chem. {\bf 102}, 9577 (1998).

\bibitem{Agam} O. Agam,
Phys. Rev. {\bf A60}, R2633 (1999); Phys. Rev. {\bf E61}, 1285 (2000).

\bibitem{Wilkinson} M. Wilkinson,
J. Phys. A {\bf 20}, 2415 (1987); J. Phys. A {\bf 21}, 1173 (1988).

\bibitem{Eckh92b} B. Eckhardt, S. Fishman, K. M\"uller and D. Wintgen,
Phys. Rev. {\bf A45} 3531 (1992).

\bibitem{HOdA} L.P. Kadanoff and C. Tang,
Proc. Natl. Acad. Sci USA {\bf 81}, 1276 (1984);
J.H. Hannay and A.M. Ozorio de Almeida, J. Phys. A {\bf 17}, 3429 (1984);
P. Cvitanovi{\'c} and B. Eckhardt, J. Phys. A {\bf 24}, L237 (1991).

\bibitem{Eckh95} B. Eckhardt, S. Fishman, J. Keating, O. Agam, J. Main
and K. M\"uller, Phys. Rev. {\bf E52}, 5893 (1995).

\bibitem{Mehlig} B. Mehlig, D. Boose and K. M\"uller,
Phys. Rev. Lett. {\bf 75}, 57 (1995);
D. Boose, J. Main, B. Mehlig and K. M\"uller,
Europhys. Lett. {\bf 32} 295 (1995);
%
%\bibitem{BK96} E.B. Bogomolny and J.P. Keating,
%Phys. Rev. Lett. {\bf 77}, 1472 (1996).
%
%\bibitem{MME} 
B. Mehlig, K. Mehlig and B. Eckhardt,
Phys. Rev. {\bf E59}, 5272 (1998).

\bibitem{Keating}
T.O. Carvalho, J.P. Keating and J.M. Robbins,
J. Phys. A, {\bf 31}, 5631 (1998).
	
\bibitem{Baecker}
A. B\"acker, R. Schubert, P. Stifter, Phys. Rev. E, {\bf 57}, 5425 (1998);
Erratum {\bf 58}, 5192 (1998)

\bibitem{FK} S. Fishman and J.P. Keating,
J. Phys. A {\bf 31}, L313 (1998).

\bibitem{Dresden} B. Eckhardt, P.~Pollner and I. Varga,
Physica E, (2000) in press

\bibitem{Eckh92} B. Eckhardt,
Chaos {\bf 3}, 613 (1993).

\bibitem{Borgonovi} F. Borgonovi, I. Guarneri, and D.L. Shepelyansky,
Phys. Rev. {\bf A43}, 4517 (1991).

\bibitem{Sommers}Y.V. Fyodorov and H.-J. Sommers 
J. Math. Phys. {\bf 38}, 1918 (1997).

\bibitem{Haake2} Haake, F. Izrailev, N. Lehmann, D. Saher, and H.-J. Sommers
Z. Phys. B {\bf 88}, 359 (1992).

\bibitem{Izraelev} F.M. Izrailev,
Phys. Rep. {\bf 196}, 299 (1990).

\bibitem{BS}R. Bl\"umel and U. Smilansky,
Phys. Rev. Lett. {\bf 69}, 217 (1992).

\bibitem{Maribor} B. Eckhardt, I. Varga and P. Pollner, Prog.
Theor. Phys. Suppl. {\bf 139}, (2000) in press.
% Proc. of
%the 4th International Summer School and Conference ``Let's Face
%Chaos through Nonlinear Dynamics'', submitted.

\bibitem{shepel}G. Casati, G. Maspero, and D.L. Shepelyansky,
Phys. Rev. Lett. {\bf 82}, 524 (1999).

\bibitem{Bogomolny} E.B. Bogomolny,
Physica(Amsterdam) {\bf 31D}, 169 (1988).

\bibitem{Huepper} B. H\"upper and B. Eckhardt,
Phys. Rev. {\bf A57}, 1536 (1998).

\bibitem{Hamburg} I. Varga, P. Pollner and B. Eckhardt,
Ann. Phys. (Leipzig) {\bf 8}, SI265 (1999).

\end{references}
\end{document}